\documentclass{llncs}
\usepackage{makeidx}  
\usepackage{latexsym}
\usepackage{graphicx}
\usepackage{algorithm2e}
\usepackage{tabularx}
\usepackage{multirow}
\usepackage{float}

\newcommand{\be}{\begin{equation}}
\newcommand{\ee}{\end{equation}}
\newcommand{\bea}{\begin{eqnarray}}
\newcommand{\eea}{\end{eqnarray}}

\newcommand{\bee}{\begin{eqnarray*}}
\newcommand{\eee}{\end{eqnarray*}}
\begin{document}
\title{\bf Implementation of Fast and Power Efficient  SEC-DAEC and SEC-DAEC-TAEC Codecs on FPGA}
\author{Sayan Tripathi \and Jhilam Jana \and Jaydeb Bhaumik\\ }
\institute{Dept. of ETCE, Jadavpur University, Kolkata, India\\
				  \email{tripathysayan@gmail.com}, \email{jhilamjana2014@gmail.com}, \email{jaydeb.bhaumik@jadavpuruniversity.in}}
\maketitle
\begin{abstract}
The reliability of memory devices is affected by radiation-induced soft errors. Multiple cell upsets (MCUs) caused by radiation corrupt data stored in multiple cells within memories. Error correction codes (ECCs) are typically used to mitigate the effects of MCUs. Single error correction-double error detection (SEC-DED) codes are not the right choice against MCUs, but are more suitable for protecting memory against single cell upset (SCU). Single error correction-double adjacent error correction (SEC-DAEC) and single error correction-double adjacent error correction-triple adjacent error correction (SEC-DAEC-TAEC) codes are more suitable due to the increasing tendency of adjacent errors. This paper presents the implementation of fast and low power multi-bit adjacent error correction codes for protecting memories. Related SEC-DAEC and SEC-DAEC-TAEC codecs with data length of 16-bit, 32-bit and 64-bit have been implemented. It is found from FPGA based implementation results that the modified designs have comparable area and have less delay and power consumption.  
\end{abstract}
\section{Introduction}
In modern high speed computing applications, static random access memory (SRAM) play a very important role as a storage subsystems. For electronic systems to operate properly, SRAM reliability is a key consideration. The main issue is the soft errors brought on by radiations, which have an impact on the SRAMs' reliability \cite{Dixit2011}. One memory cell is corrupted by a soft error in a single cell upset (SCU). However the multiple cell upsets (MCUs) are a prevalent event with downscaling of technology nodes \cite{Ibe2010}. ECCs are mostly used in memories as a defence against these soft errors. In the past, SEC codes were mainly useful for protecting SRAMs from SCUs \cite{Hamming1950}-\cite{Hsiao1970}. To protect memories against MCUs, interleaved SEC-DED codes have been used, but they are more complex. Recently, DAEC \cite{Dutta2007}-\cite{Tripathi2023}, TAEC \cite{Neale2015}-\cite{Moran2018} and burst error correction (BEC) codes \cite{Li20191}-\cite{Li20181} have been introduced. \\
Neale et al. have presented a technique for designing SEC-DED-DAEC codes that has the additional capability of scaling adjacent error detection (xAED) \cite{Neale2013}-\cite{Neale2014}. Further modifications were made to these codes in order to implement the TAEC feature \cite{Neale2015}. A method for double adjacent ECCs that has zero miscorrection for memories was proposed by Dutta et al. \cite{Dutta2012}. Reviriego et al. have presented area and delay optimisation strategies for SEC-DED-DAEC codes \cite{Reviriego2014}. Neale et al. \cite{Neale2015} and Adalid et al. \cite{Adalid2014} developed the triple adjacent error correcting codes. But these codes need complicated decoding circuitry and have higher delay which make their design more gate-intensive. Tripathi et al. have implemented an efficient multi-bit adjacent ECC for memory \cite{Tripathi2021}. Also, Moran et al. presented a flexible unequal error correction codes. \cite{Moran2018}. Li et al. proposed 3-bit BEC codes which have lower encoder and decoder delay \cite{Li20191}. \\
In this paper, we have introduced modified SEC-DAEC and SEC-DAEC-TAEC codes with 16, 32, and 64 information bits and 0\% miscorrection rate in order to increase the reliability of storage systems against soft errors and make the system delay and power efficient. On the FPGA platform, performance of modified and existing codes has been analysed in terms of area, delay and power. \\
The remaining part of our paper is structured as follows. Section 2 provides basics of SEC-DAEC and SEC-DAEC-TAEC codes. Section 3 describes the corrections on existing SEC-DAEC and SEC-DAEC-TAEC codes. Section 4 presents implementation results and finally concluding remarks are made in Section 5.
\section{Basics of SEC-DAEC and SEC-DAEC-TAEC Codes}
In this section, basic overview of $H$-matrix construction for SEC-DAEC and SEC-DAEC-TAEC codes is presented. The $H$-matrices for SEC-DAEC and SEC-DAEC-TAEC codes are constructed using some design constraints which are as follows: i) every columns should be non-zero and have unique values, (ii) XOR sum of any two adjacent columns must be unique and non-zero, should not be equal to any of the individual column, (iii) XOR sum of any three adjacent columns must be unequal to any of the individual column and non-zero. The first condition confirm the SEC capability. DAEC property is satisfied by conditions (i) and (ii). TAEC property is confirmed by constraints (i), (ii) and (iii). \\

\section{Corrections on Existing SEC-DAEC and SEC-DAEC-TAEC Codes [12]}
In this section, some corrections on existing SEC-DAEC and SEC-DAEC-TAEC codes \cite{Tripathi2023} are presented.  For the sake of completeness, encoding and decoding processes of (23 16) SEC-DAEC-TAEC code \cite{Tripathi2023} is described here. The (23, 16) $H$-matrix \cite{Tripathi2023} is shown in Fig. \ref{fig10000}. 
\begin{figure}[]
\centering
\newcommand\scalemath[2]{\scalebox{#1}{\mbox{\ensuremath{\displaystyle #2}}}}
\[
H=
\left[
	\scalemath{1.2}{
\begin{array}{cccccccccccccccccccccccc}
0	0	0	0	0	1	0	1	1	0	1	0	1	0	1	0	0	0	0	0	1	1	1\\
0	1	0	1	1	0	1	0	0	0	0	0	0	1	0	1	0	0	0	0	1	1	1\\
0	1	0	1	0	0	0	0	0	1	0	1	1	0	1	0	1	0	1	0	0	0	0\\
1	0	0	0	1	0	0	0	1	0	0	0	1	0	0	0	1	0	0	0	1	0	0\\
0	1	0	0	0	1	0	0	0	1	0	0	0	1	0	0	0	1	0	0	0	1	0\\
0	0	1	0	0	0	1	0	0	0	1	0	0	0	1	0	0	0	1	0	0	0	1\\
0	0	0	1	0	0	0	1	0	0	0	1	0	0	0	1	0	0	0	1	0	0	0\\
 \end{array}
}
\right]
\]
\caption{$H$-matrix of SEC-DAEC and SEC-DAEC-TAEC (23, 16) code}
\label{fig10000}
\end{figure}
Parity bits are calculated from information bits during encoding, and the associated codeword is stored in the memory.
Parity equations associated with (23, 16) $H$-matrix are provided in equation (\ref{equ2}).
\begin{eqnarray}
\label{equ2}
p_{b1} & =  & i_{b3} \oplus i_{b10} \oplus i_{b13} \oplus i_{b14} \oplus p_{b3} \nonumber\\
p_{b2} & =  & i_{b5} \oplus i_{b8} \oplus i_{b11} \oplus i_{b16}\oplus p_{b6} \nonumber\\
p_{b3} & =  & i_{b4} \oplus i_{b6} \oplus i_{b8} \oplus i_{b10} \oplus i_{b11} \oplus i_{b14}\oplus i_{b15}\oplus i_{b16} \nonumber\\
p_{b4} & = & i_{b1} \oplus i_{b2} \oplus i_{b3} \oplus i_{b5} \oplus i_{b12} \oplus i_{b14} \oplus i_{b15} \oplus i_{b16} \nonumber\\
p_{b5} & = & i_{b1} \oplus i_{b4} \oplus i_{b7} \oplus i_{b15} \oplus p_{b4} \nonumber\\
p_{b6} & = & i_{b1} \oplus i_{b2} \oplus i_{b7} \oplus i_{b9} \oplus i_{b10} \oplus i_{b11} \oplus i_{b13} \nonumber\\
p_{b7} & = & i_{b2} \oplus i_{b6} \oplus i_{b9} \oplus i_{b12}  \nonumber\\
\end{eqnarray}
During the decoding process, the syndrome is generated by multiplying the received codeword with the transpose of $H$-matrix. Syndrome values are calculated to locate the error in stored codeword. In the encoding process of section 3 \cite{Tripathi2023}, for 16-bit information word $i_b$=$1111111111111111$, the encoder computes seven parity bits 0100010. Therefore, generated codeword will be \textbf{01111111011110111\\010111} which is obtained by appending parity bits with information bits and stored in the memory. To demonstrate the correction capability of code, errors are injected on three adjacent bits on $i_{b2}$, $i_{b3}$ and $i_{b4}$ in the generated codeword. After injection of errors, the received codeword will be \textbf{011\underline{000}11011110111010\\111}. The error in received codeword are corrected using syndrome bits (1011101). Finally, detected error is corrected using error correction logic.\\
In article \cite{Tripathi2023}, there are some mistakes in Fig. 4 like the value of the codeword stored in memory cells, injected errors, syndrome bits, codeword read from memory and codeword after error correction. The rectified figure has been depicted here as Fig. \ref{fig1}.
\begin{figure}[] 
\centering
\includegraphics[width=5.0in]{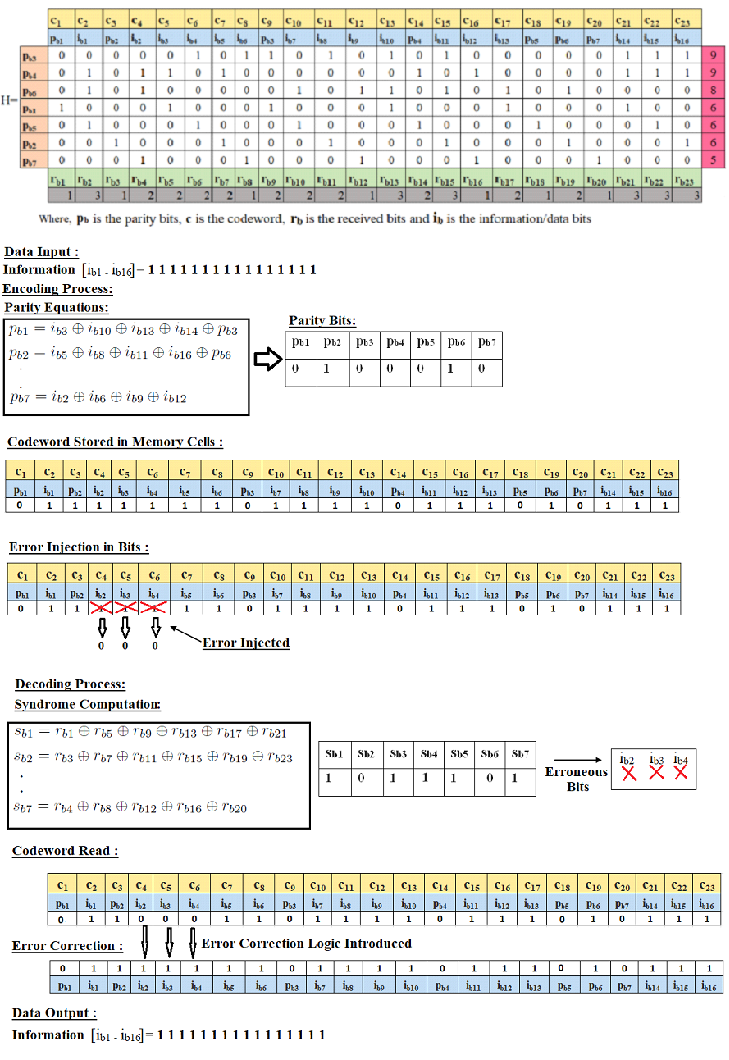}
\caption{Encoding and decoding methodologies}
\label{fig1}
\end{figure}
Also, Fig. 5 in \cite{Tripathi2023} has been corrected and presented in Fig. \ref{fig2}.
\begin{figure}[]
\centering
\includegraphics[width=5.2in]{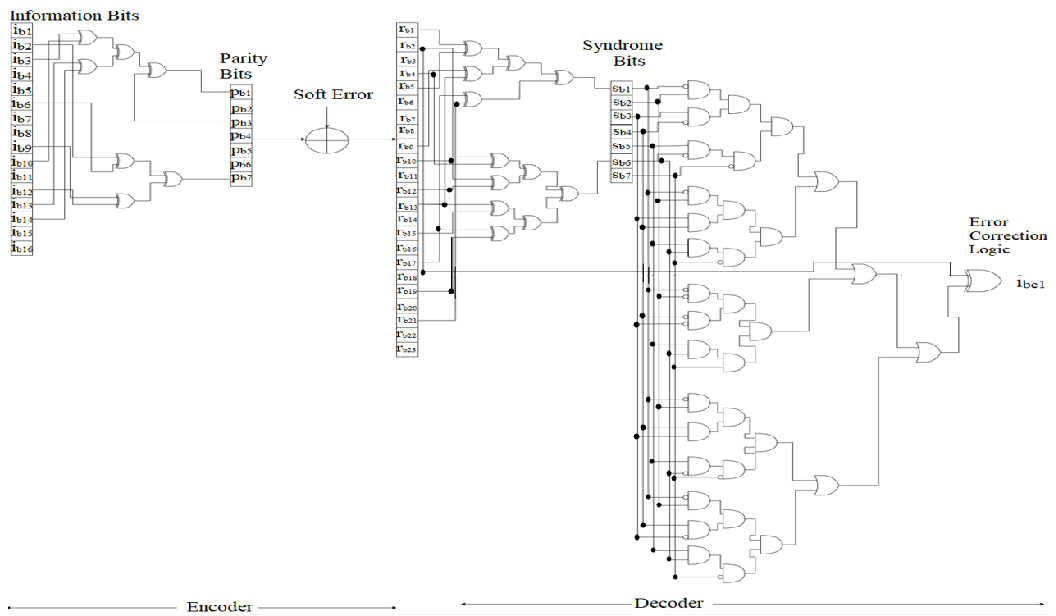}
\caption{Gate level design of modified (23, 16) SEC-DAEC-TAEC codec}
\label{fig2}
\end{figure}

\section{FPGA-Based Implementation Results}
In this section, FPGA-based implementation of (23, 16), (40, 32) and (74, 64) SEC-DAEC and SEC-DAEC-TAEC codecs are presented. Several SEC-DAEC and SEC-DAEC-TAEC codes have been described in Verilog and implemented in FPGA platform. The modified and existing codes are implemented using Zynq UltraScale+ MPSoC (ZCU104) FPGA evaluaton kit. The performances of all designs are observed with respect to look-up tables (LUTs), delay, power for FPGA-implementation which are presented in Table \ref{tabsyn}. The performance of the modified SEC-DAEC and SEC-DAEC-TAEC codecs have been obtained by using common sub expression. The highest improvement in area (LUTs) is 47.03\% and 36.54\% achieved for the modified implementation of SEC-DAEC code and SEC-DAEC-TAEC code respectively. Also the highest 17.95\%  delay improvement for implemented SEC-DAEC code and SEC-DAEC-TAEC code is obtained against Moran et al.\cite{Moran2018}. The highest improvement in power is 21.62\% and 22.30\% achieved in FPGA-based implementation of SEC-DAEC code and SEC-DAEC-TAEC code against Moran et al.\cite{Moran2018} and Neale et al. 
 \cite{Neale2015} respectively.
\begin{table}[H]
\caption{FPGA implementation results of modified and existing codecs}
\label{tabsyn}
\centering
\resizebox{!}{0.58\textheight}{%
\begin{tabular}{|c|c|c|c|c|c|c|c|c|}
\hline
\multirow{2}{*}{Scheme} & \multirow{2}{*}{\begin{tabular}[c]{@{}c@{}}Data \\ Bits\end{tabular}} & \multirow{2}{*}{Codec} & \multirow{2}{*}{\begin{tabular}[c]{@{}c@{}}Area \\ (LUTs)\end{tabular}} & \multirow{2}{*}{\begin{tabular}[c]{@{}c@{}}Delay \\ (ns)\end{tabular}} & \multirow{2}{*}{\begin{tabular}[c]{@{}c@{}}Power \\ (W)\end{tabular}} & \multirow{2}{*}{\begin{tabular}[c]{@{}c@{}}Impro. in \\ Area (\%)\end{tabular}} & \multirow{2}{*}{\begin{tabular}[c]{@{}c@{}}Impro. in \\ Delay (\%)\end{tabular}} & \multirow{2}{*}{\begin{tabular}[c]{@{}c@{}}Impro. in \\ Power (\%)\end{tabular}} \\
 &  &  &  &  &  &  &  &  \\ \hline
\multirow{16}{*}{DAEC} & \multirow{6}{*}{16} & \begin{tabular}[c]{@{}c@{}}Neale et al.  \\ (23, 16) \cite{Neale2014}\end{tabular} & 68 & 3.24 & 2.89 & 20.59 & 3.70 & 19.72 \\ \cline{3-9} 
 &  & \begin{tabular}[c]{@{}c@{}}Reviriego et al. \\ (23, 16) \cite{Reviriego2014}\end{tabular} & 65 & 3.18 & 2.72 & 16.92 & 1.89 & 14.71 \\ \cline{3-9} 
 &  & \begin{tabular}[c]{@{}c@{}}Moran et al. \\ (23, 16) \cite{Moran2018}\end{tabular} & 72 & 3.49 & 2.96 & 25.00 & \textbf{10.60} & \textbf{21.62} \\ \cline{3-9} 
 &  & \begin{tabular}[c]{@{}c@{}}Dutta et al. \\ (23, 16) \cite{Dutta2012}\end{tabular} & 66 & 3.22 & 2.86 & 18.18 & 3.11 & 18.88 \\ \cline{3-9} 
 &  & \begin{tabular}[c]{@{}c@{}}Tripathi et al. \\ (23, 16) \cite{Tripathi2023}\end{tabular} & 64 & 3.16 & 2.34 & 15.63 & 1.27 & 0.85 \\ \cline{3-9} 
 &  & \begin{tabular}[c]{@{}c@{}}Modified \\ (23, 16)\end{tabular} & 54 & 3.12 & 2.32 & - & - & - \\ \cline{2-9} 
 & \multirow{5}{*}{32} & \begin{tabular}[c]{@{}c@{}}Neale et al. \\ (40, 32) \cite{Neale2014}\end{tabular} & 185 & 4.37 & 3.95 & \textbf{47.03} & 8.47 & 7.85 \\ \cline{3-9} 
 &  & \begin{tabular}[c]{@{}c@{}}Reviriego et al. \\ (39, 32) \cite{Reviriego2014}\end{tabular} & 114 & 4.06 & 3.83 & 14.04 & 1.48 & 4.96 \\ \cline{3-9} 
 &  & \begin{tabular}[c]{@{}c@{}}Dutta et al. \\ (40, 32) \cite{Dutta2012}\end{tabular} & 139 & 4.31 & 4.07 & 29.50 & 7.19 & 10.57 \\ \cline{3-9} 
 &  & \begin{tabular}[c]{@{}c@{}}Tripathi et al. \\ (40, 32) \cite{Tripathi2023}\end{tabular} & 120 & 4.03 & 3.67 & 18.33 & 0.74 & 0.82 \\ \cline{3-9} 
 &  & \begin{tabular}[c]{@{}c@{}}Modified \\ (40, 32)\end{tabular} & 98 & 4.00 & 3.64 & - & - & - \\ \cline{2-9} 
 & \multirow{5}{*}{64} & \begin{tabular}[c]{@{}c@{}}Neale et al.  \\ (74, 64) \cite{Neale2014}\end{tabular} & 261 & 4.61 & 4.64 & 22.22 & 6.94 & 6.47 \\ \cline{3-9} 
 &  & \begin{tabular}[c]{@{}c@{}}Reviriego et al. \\ (73, 64) \cite{Reviriego2014}\end{tabular} & 214 & 4.53 & 4.44 & 5.14 & 5.30 & 2.25 \\ \cline{3-9} 
 &  & \begin{tabular}[c]{@{}c@{}}Dutta et al. \\ (73, 64) \cite{Dutta2012}\end{tabular} & 278 & 4.46 & 4.76 & 26.98 & 3.81 & 8.82 \\ \cline{3-9} 
 &  & \begin{tabular}[c]{@{}c@{}}Tripathi et al. \\ (74, 64) \cite{Tripathi2023}\end{tabular} & 253 & 4.41 & 4.40 & 19.76 & 2.72 & 1.36 \\ \cline{3-9} 
 &  & \begin{tabular}[c]{@{}c@{}}Modified \\ (74, 64)\end{tabular} & 203 & 4.29 & 4.34 & - & - & - \\ \hline
\multirow{16}{*}{TAEC} & \multirow{6}{*}{16} & \begin{tabular}[c]{@{}c@{}}Neale et al. \\ (23, 16) \cite{Neale2015}\end{tabular} & 78 & 3.71 & 3.70 & 15.38 & 13.75 & 21.08 \\ \cline{3-9} 
 &  & \begin{tabular}[c]{@{}c@{}}Adalid et al. \\ (22, 16) \cite{Adalid2014}\end{tabular} & 58 & 3.31 & 3.42 & -13.79 & 3.32 & 14.62 \\ \cline{3-9} 
 &  & \begin{tabular}[c]{@{}c@{}}Moran et al. \\ (24, 16) \cite{Moran2018}\end{tabular} & 104 & 3.90 & 3.52 & \textbf{36.54} & \textbf{17.95} & 17.05 \\ \cline{3-9} 
 &  & \begin{tabular}[c]{@{}c@{}}Li et al. \\ (23, 16) \cite{Li20191}\end{tabular} & 55 & 3.36 & 3.00 & -20.00 & 4.76 & 2.67 \\ \cline{3-9} 
 &  & \begin{tabular}[c]{@{}c@{}}Tripathi et al. \\ (23, 16) \cite{Tripathi2023}\end{tabular} & 76 & 3.26 & 2.94 & 13.16 & 1.84 & 0.68 \\ \cline{3-9} 
 &  & \begin{tabular}[c]{@{}c@{}}Modified \\ (23, 16)\end{tabular} & 66 & 3.20 & 2.92 & - & - & - \\ \cline{2-9} 
 & \multirow{5}{*}{32} & \begin{tabular}[c]{@{}c@{}}Neale et al. \\ (40, 32) \cite{Neale2015}\end{tabular} & 177 & 3.89 & 5.92 & 18.64 & 4.37 & \textbf{22.30} \\ \cline{3-9} 
 &  & \begin{tabular}[c]{@{}c@{}}Adalid et al. \\ (39, 32) \cite{Adalid2014}\end{tabular} & 166 & 4.32 & 5.78 & 13.25 & 13.89 & 20.42 \\ \cline{3-9} 
 &  & \begin{tabular}[c]{@{}c@{}}Li et al. \\ (40, 32) \cite{Li20191}\end{tabular} & 160 & 4.06 & 5.12 & 10.00 & 8.37 & 10.16 \\ \cline{3-9} 
 &  & \begin{tabular}[c]{@{}c@{}}Tripathi et al. \\ (40, 32) \cite{Tripathi2023}\end{tabular} & 170 & 3.78 & 4.64 & 15.29 & 1.59 & 0.86 \\ \cline{3-9} 
 &  & \begin{tabular}[c]{@{}c@{}}Modified \\ (40, 32)\end{tabular} & 144 & 3.72 & 4.60 & - & - & - \\ \cline{2-9} 
 & \multirow{5}{*}{64} & \begin{tabular}[c]{@{}c@{}}Neale et al. \\ (74, 64) \cite{Neale2015}\end{tabular} & 374 & 4.79 & 6.36 & 16.58 & 9.81 & 18.87 \\ \cline{3-9} 
 &  & \begin{tabular}[c]{@{}c@{}}Adalid et al. \\ (72, 64) \cite{Adalid2014}\end{tabular} & 329 & 4.51 & 6.26 & 5.17 & 4.21 & 17.57 \\ \cline{3-9} 
 &  & \begin{tabular}[c]{@{}c@{}}Li et al. \\ (73, 64) \cite{Li20191}\end{tabular} & 389 & 4.60 & 5.84 & 19.79 & 6.09 & 11.64 \\ \cline{3-9} 
 &  & \begin{tabular}[c]{@{}c@{}}Tripathi et al. \\ (74, 64) \cite{Tripathi2023} \end{tabular} & 362 & 4.41 & 5.20 & 13.81 & 2.04 & 0.77 \\ \cline{3-9} 
 &  & \begin{tabular}[c]{@{}c@{}}Modified \\ (74, 64)\end{tabular} & 312 & 4.32 & 5.16 & - & - & - \\ \hline
\end{tabular}
}
\end{table}
Table \ref{tabsyn1} represents the performance of modified and existing ECCs in terms of LUTs-delay product (LDP), power-LUTs Product (PLP), power-delay product (PDP) in FPGA platform. The highest improvement in terms of LDP, PLP are obtained compared to Neale et al. scheme and the highest improvement in terms of PDP is achieved against Moran et al. scheme .
\begin{table}[]
\caption{Improvement calculation of FPGA-based implementation results}
\label{tabsyn1}
\centering
\resizebox{!}{0.5\textheight}{%
\begin{tabular}{|c|c|c|c|c|c|c|c|c|}
\hline
\multirow{2}{*}{Scheme} & \multirow{2}{*}{\begin{tabular}[c]{@{}c@{}}Data \\ Bits\end{tabular}} & \multirow{2}{*}{Codec} & \multirow{2}{*}{\begin{tabular}[c]{@{}c@{}}LUTs Delay \\ Product (LDP)\end{tabular}} & \multirow{2}{*}{\begin{tabular}[c]{@{}c@{}}Power LUTs \\ Product (PLP)\end{tabular}} & \multirow{2}{*}{\begin{tabular}[c]{@{}c@{}}Power Delay \\ Product (PDP)\end{tabular}} & \multirow{2}{*}{\begin{tabular}[c]{@{}c@{}}Impro. in \\ LDP (\%)\end{tabular}} & \multirow{2}{*}{\begin{tabular}[c]{@{}c@{}}Impro. in \\ PLP (\%)\end{tabular}} & \multirow{2}{*}{\begin{tabular}[c]{@{}c@{}}Impro. in \\ PDP (\%)\end{tabular}} \\
 &  &  &  &  &  &  &  &  \\ \hline
\multirow{16}{*}{DAEC} & \multirow{6}{*}{16} & \begin{tabular}[c]{@{}c@{}}Neale et al.  \\ (23, 16) \cite{Neale2014}\end{tabular} & 220.32 & 196.52 & 9.36 & 23.53 & 36.25 & 22.70 \\ \cline{3-9} 
 &  & \begin{tabular}[c]{@{}c@{}}Reviriego et al. \\ (23, 16) \cite{Reviriego2014}\end{tabular} & 206.70 & 176.80 & 8.65 & 18.49 & 29.14 & 16.32 \\ \cline{3-9} 
 &  & \begin{tabular}[c]{@{}c@{}}Moran et al. \\ (23, 16) \cite{Moran2018}\end{tabular} & 251.28 & 213.12 & 10.33 & 32.95 & 41.22 & 29.93 \\ \cline{3-9} 
 &  & \begin{tabular}[c]{@{}c@{}}Dutta et al. \\ (23, 16) \cite{Dutta2012}\end{tabular} & 212.52 & 188.76 & 9.21 & 20.72 & 33.63 & 21.40 \\ \cline{3-9} 
 &  & \begin{tabular}[c]{@{}c@{}}Tripathi et al. \\ (23, 16) \cite{Tripathi2023}\end{tabular} & 202.24 & 149.76 & 7.39 & 16.69 & 16.35 & 2.11 \\ \cline{3-9} 
 &  & \begin{tabular}[c]{@{}c@{}}Modified \\ (23, 16)\end{tabular} & 168.48 & 125.28 & 7.24 & - & - & - \\ \cline{2-9} 
 & \multirow{5}{*}{32} & \begin{tabular}[c]{@{}c@{}}Neale et al.  \\ (40, 32) \cite{Neale2014}\end{tabular} & 808.45 & 730.75 & 17.26 & 51.51 & 51.18 & 15.65 \\ \cline{3-9} 
 &  & \begin{tabular}[c]{@{}c@{}}Reviriego et al. \\ (39, 32) \cite{Reviriego2014}\end{tabular} & 462.84 & 436.62 & 15.55 & 15.31 & 18.30 & 6.37 \\ \cline{3-9} 
 &  & \begin{tabular}[c]{@{}c@{}}Dutta et al. \\ (40, 32) \cite{Dutta2012}\end{tabular} & 599.09 & 565.73 & 17.54 & 34.57 & 36.95 & 17.00 \\ \cline{3-9} 
 &  & \begin{tabular}[c]{@{}c@{}}Tripathi et al. \\ (40, 32) \cite{Tripath2023}\end{tabular} & 483.60 & 440.40 & 14.79 & 18.94 & 19.00 & 1.56 \\ \cline{3-9} 
 &  & \begin{tabular}[c]{@{}c@{}}Modified \\ (40, 32)\end{tabular} & 392.00 & 356.72 & 14.56 & - & - & - \\ \cline{2-9} 
 & \multirow{5}{*}{64} & \begin{tabular}[c]{@{}c@{}}Neale et al.  \\ (74, 64) \cite{Neale2014}\end{tabular} & 1203.21 & 1211.04 & 21.39 & 27.62 & 27.25 & 12.96 \\ \cline{3-9} 
 &  & \begin{tabular}[c]{@{}c@{}}Reviriego et al. \\ (73, 64) \cite{Reviriego2014}\end{tabular} & 969.42 & 950.16 & 20.11 & 10.17 & 7.28 & 7.43 \\ \cline{3-9} 
 &  & \begin{tabular}[c]{@{}c@{}}Dutta et al. \\ (73, 64) \cite{Dutta2012}\end{tabular} & 1239.88 & 1323.28 & 21.23 & 29.76 & 33.42 & 12.30 \\ \cline{3-9} 
 &  & \begin{tabular}[c]{@{}c@{}}Tripathi et al. \\ (74, 64) \cite{Tripathi2023}\end{tabular} & 1115.73 & 1113.20 & 19.40 & 21.95 & 20.86 & 4.05 \\ \cline{3-9} 
 &  & \begin{tabular}[c]{@{}c@{}}Modified \\ (74, 64)\end{tabular} & 870.87 & 881.02 & 18.62 & - & - & - \\ \hline
\multirow{16}{*}{TAEC} & \multirow{6}{*}{16} & \begin{tabular}[c]{@{}c@{}}Neale et al. \\ (23, 16) \cite{Neale2015}\end{tabular} & 289.38 & 288.60 & 13.73 & 27.02 & 33.22 & 31.93 \\ \cline{3-9} 
 &  & \begin{tabular}[c]{@{}c@{}}Adalid et al. \\ (22, 16) \cite{Adalid2014}\end{tabular} & 191.98 & 198.36 & 11.32 & -10.01 & 2.84 & 17.46 \\ \cline{3-9} 
 &  & \begin{tabular}[c]{@{}c@{}}Moran et al. \\ (24, 16) \cite{Moran2018}\end{tabular} & 405.60 & 366.08 & 13.73 & 47.93 & 47.36 & 31.93 \\ \cline{3-9} 
 &  & \begin{tabular}[c]{@{}c@{}}Li et al. \\ (23, 16) \cite{Li20191}\end{tabular} & 184.80 & 165.00 & 10.08 & -14.29 & -16.80 & 7.30 \\ \cline{3-9} 
 &  & \begin{tabular}[c]{@{}c@{}}Tripathi et al. \\ (23, 16) \cite{Tripathi2023}\end{tabular} & 247.76 & 223.44 & 9.58 & 14.76 & 13.75 & 2.51 \\ \cline{3-9} 
 &  & \begin{tabular}[c]{@{}c@{}}Modified \\ (23, 16)\end{tabular} & 211.20 & 192.72 & 9.34 & - & - & - \\ \cline{2-9} 
 & \multirow{5}{*}{32} & \begin{tabular}[c]{@{}c@{}}Neale et al. \\ (40, 32) \cite{Neale2015}\end{tabular} & 688.53 & 1047.84 & 23.03 & 22.20 & 36.78 & 25.69 \\ \cline{3-9} 
 &  & \begin{tabular}[c]{@{}c@{}}Adalid et al. \\ (39, 32) \cite{Adalid2014}\end{tabular} & 717.12 & 959.48 & 24.97 & 25.30 & 30.96 & 31.47 \\ \cline{3-9} 
 &  & \begin{tabular}[c]{@{}c@{}}Li et al. \\ (40, 32) \cite{Li20191}\end{tabular} & 649.60 & 819.20 & 20.79 & 17.54 & 19.14 & 17.68 \\ \cline{3-9} 
 &  & \begin{tabular}[c]{@{}c@{}}Tripathi et al. \\ (40, 32) \cite{Tripathi2023}\end{tabular} & 642.60 & 788.80 & 17.54 & 16.64 & 16.02 & 2.44 \\ \cline{3-9} 
 &  & \begin{tabular}[c]{@{}c@{}}Modified \\ (40, 32)\end{tabular} & 535.68 & 662.40 & 17.11 & - & - & - \\ \cline{2-9} 
 & \multirow{5}{*}{64} & \begin{tabular}[c]{@{}c@{}}Neale et al. \\ (74, 64) \cite{Neale2015}\end{tabular} & 1791.46 & 2378.64 & 30.46 & 24.76 & 32.32 & 26.83 \\ \cline{3-9} 
 &  & \begin{tabular}[c]{@{}c@{}}Adalid et al. \\ (72, 64) \cite{Adalid2014}\end{tabular} & 1483.79 & 2059.54 & 28.23 & 9.16 & 21.83 & 21.04 \\ \cline{3-9} 
 &  & \begin{tabular}[c]{@{}c@{}}Li et al. \\ (73, 64) \cite{Li20191}\end{tabular} & 1789.40 & 2271.76 & 26.86 & 24.68 & 29.13 & 17.02 \\ \cline{3-9} 
 &  & \begin{tabular}[c]{@{}c@{}}Tripathi et al. \\ (74, 64) \cite{Tripathi2023}\end{tabular} & 1596.42 & 1882.40 & 22.93 & 15.57 & 14.48 & 2.79 \\ \cline{3-9} 
 &  & \begin{tabular}[c]{@{}c@{}}Modified \\ (74, 64)\end{tabular} & 1347.84 & 1609.92 & 22.29 & - & - & - \\ \hline
\end{tabular}
}
\end{table}
\section{Conclusion}
In this article, modified FPGA-based implementation of fast and power efficient SEC-DAEC and SEC-DAEC-TAEC codes with 0\% miscorrection are presented. On FPGA platform, modified and other related codes have been implemented. The highest improvement of 47.03\% in area, 17.95\% in delay and 22.30\% in power have been achieved in FPGA implementation. The results show that the modified implementation has low delay and are power efficient than existing designs. Consequently, our implemented SEC-DAEC and SEC-DAEC-TAEC codes can be used in memory subsystem.

\end{document}